\documentclass[aps,pre,twocolumn,showpacs,
nolongbibliography,10pt]{revtex4-2}

\usepackage[sort&compress]{natbib}
\usepackage{scalerel}
\usepackage[normalem]{ulem}
\usepackage{xcolor}
\usepackage{bm}
\usepackage{amsmath}
\usepackage{amsfonts}
\usepackage{amssymb}
\usepackage{graphicx}
\usepackage{subfigure}
\usepackage{mathtools}
\usepackage{physics}
\usepackage{dsfont}
\usepackage{float}
\usepackage{lmodern}
\usepackage[many]{tcolorbox}
\usepackage{empheq}
\usepackage{cleveref}
\usepackage{textcomp}
\newcommand{\scs}{\scriptscriptstyle}

\begin{document}

\title{Transient chaos and Rayleigh particle escape out of a time modulated optical trap}
\author{Evgeny N. Bulgakov$^1$}
\author{Konstantin N. Pichugin$^1$}
\author{Dmitrii N. Maksimov$^{1,2}$}
\affiliation{$^1$LV Kirensky Institute of Physics, Federal Research Centre KSC SB RAS, 660036, Krasnoyarsk, Russia}
\affiliation{$^2$IRC SQC, Siberian Federal University, 660041, Krasnoyarsk, Russia}

\date{\today}

\begin{abstract}
We consider Rayleigh particles in a periodically modulated optical trap formed by two counter-propagating Gaussian beams. It is shown that for certain values of parameters the system exhibits transient chaos which manifests itself in particle acceleration and subsequent directional ejection out of the trap. The escape flights are terminated at the distance of hundreds wavelengths from the trap centrum and the particles return to the trap under the action of the Stokes force. The particle escape is shown to be a threshold effect that can be potentially employed for particle sorting.  
\end{abstract}
\maketitle


{\it Introduction} --
Optomechanical systems have attracted significant interest due to their broad utility in hybrid transduction platforms, precision sensing, quantum measurement technologies, and optical sorting \cite{Aspelmeyer2014, Toftul2024a, yang2025optical}. 
Optomechanical systems exhibit a wide variety of nonlinear phenomena including dynamical multistability 
\cite{Marquardt2006, ludwig2008optomechanical, lorch2014laser, Ma2020, fu2022optomechanically},  chaos \cite{Carmon2007, bakemeier2015route, NavarroUrrios2017, djorwe2018frequency, roque2020nonlinear, Zhu2023, Zhang2024c, Saiko2025}, 
frequency comb generation \cite{Hu2021, Liu2024},
and solitons \cite{Zhang2021a}. Recently, non-linear effects were thoroughly studied in different cavity optomechanical systems where an optical bosonic mode is coupled with a mechanical \cite{Zhang2024b, Das2023, Xu2024, bibak2023dissipative, Liang2024} and a magnon mode \cite{Shen2022, Li2023a}. In the field of classical physics, nonlinear effects, such as Hopf bifurcation to a limit cycle attractor \cite{Svak2018, Simpson2021}, can be observed with optical tweezers which make it possible to manipulate sub-micron particles by applying optical forces induced by focused beams of light \cite{Polimeno2018, Pesce2020}.  

In this work, we investigate optomechanical chaos with spherical Rayleigh particles subject to the electromagnetic field of two counter-propagating Gaussian beams, one of which is modulated in time.
The longitudinal dynamics is shown to be asymptotically controlled by a "wave-like"  potential \cite{hanggi2005brownian} that is ubiquitous in wave-related transport set-ups including 
particle transport in plasma \cite{friedland2006autoresonant}, optical lattices \cite{schiavoni2003phase}, meandering flows \cite{budyansky2009detection}, and semiconductor superlattices \cite{Greenaway2010}. The action of such time-modulated potentials results in the effect of rocked ratchets \cite{schiavoni2003phase, carlo2006chaotic, Machura2005, Gommers2008, arzola2011experimental, jakl2014optical, dupont2023hamiltonian} that manifests itself in a directional motion of a particle, which initially has zero momentum.
Space-periodic potentials perturbed by time-periodic modulation are  theoretically shown to exhibit directed transport both in stochastic layers and regular islands \cite{flach2000directed, makarov2010frequency}. In the case of non-periodic potentials it is demonstrated in \cite{hennig2010transient} that 
the transient chaotic dynamics can be terminated by a regular net directed motion. 

\begin{figure}[t!]
\includegraphics[width=0.5\textwidth,height=.14\textwidth,trim=6.5cm 8.8cm 3.5cm 7.8cm,clip]{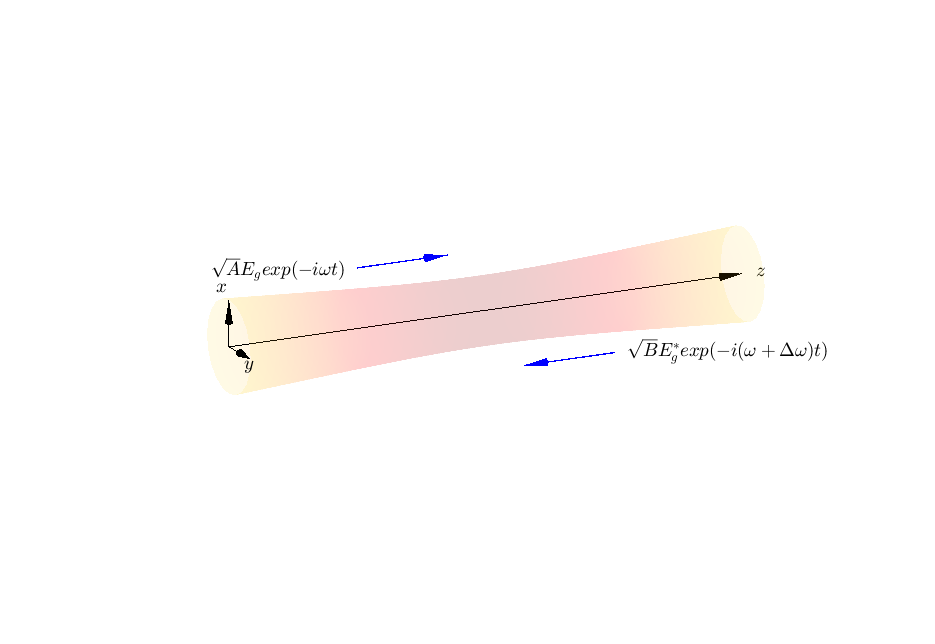}
\caption{Periodically modulated optical formed by two counter-propagating beams with the same polarization but different intensities.}
\label{fig1}
\end{figure}

According to \cite{novotny2012principles} the expression for the optical field of a Gaussian beam reads
\begin{equation}\label{gauss}
E_g(\rho,z)\!=\!\frac{4}{ \sqrt{c w_0^2}}\frac{1}{1+i{z}/{z_0}}
e^{
ik z-\frac{\rho^2}{\left(1+i{z}/{z_0}\right)w_0^2} 
},
\end{equation}
where $x,y,z$ are the Cartesian coordinates with $z$ being the beam axis, $k$ is the wavenumber, $c$ is the speed of light, 
$\rho^2={x^2+y^2}$, $z_0=kw_0^2/2$, and $w_0$ is the beam waist.  In what follows, we consider a coherent sum of two counter-propagating beams with the same $x$-polarization and the same focal plane
\begin{equation}\label{trap}
E_x =\sqrt{A} E_g(\rho,z)e^{-i\omega t}+ \sqrt{B} e^{-i(\omega+\Delta\omega) t}E_g^*(\rho,z),
\end{equation}
where the parameters $A$ and $B$ are the powers of the counter-propagating beams, $\omega=ck$ is the frequency, and $\Delta\omega\ll\omega$ accounts for the linear modulation of the beam. The schematic of the set-up is sketched in Fig.~\ref{fig1}.

{\it Hamiltonian dynamics} -- There are two optical forces acting on a Rayleigh particle in an optical trap, namely the scattering and the gradient forces \cite{novotny2012principles}. In this work, we consider Rayleigh particles made of lossless dielectric and, therefore, ignore the scattering force. The optical gradient force is expressed via the optical field as follows
\begin{equation}\label{force}
{\bf F}=\frac{1}{4}\alpha\nabla |E_x|^2,
\end{equation}
where the polarizability $\alpha$ is given by the Clausius–Mossotti relation
$
\alpha= R^3 ({\epsilon-1})/({\epsilon+2}
)$
with $R$ as the radius of the particle and $\epsilon$ as the relative permittivity of the particle. We mention in passing that Eq.~\eqref{force} is derived in \cite{novotny2012principles} as the average per cycle under the assumption of monochromatic field. In our case, this derivation applies since $\Delta\omega\ll\omega$ and the time modulation is vanishingly small on the time scale determined by $\omega$.
In air or vacuum, the optical gradient force attracts particles to the hot-spot of the optical trap. For the start, we assume that the particles are in a stable equilibrium on the $z$-axis, so the lateral motion is suppressed, $x,y=0$. Then, the longitudinal dynamics is controlled by Newton's second law of motion
\begin{equation}\label{Newton}
\frac{d^2 q}{d\tau^2}=-\frac{\partial V(q,\tau)}{\partial q},
\end{equation}
where we introduced a 1D optical potential 
\begin{align}\label{potential}
& V(q,\tau)=V_s(q)+V_{d}(q,\tau),
\end{align}
that is a sum of the static $V_s(q)$ and the dynamic contributions
$V_d(q,\tau)$
\begin{align}\label{potential2}
& V_s(q)=-\frac{1}{1+\left( {q}/{q_0}\right)^2}, \nonumber \\
& \!V_d(q,\tau)\!=\!-\delta\!\left(\!\frac{\cos(4\pi q-\bar{\omega} \tau)}{1+\left( {q}/{q_0}\right)^2} 
\!+\!2\frac{q}{q_0}\frac{\sin(4\pi q-\bar{\omega} \tau)}{\left[1+\left( {q}/{q_0}\right)^2\right]^2}\!\right)
\end{align}
where we used the following variable and parameter change
\begin{align}\label{vars}
& q=\frac{z}{\lambda_0}, \ \tau=\sqrt{\frac{4\alpha(A\!+\!B)}{ m c \lambda_0^2 w_0^2}}t, \ p=\frac{dq}{d\tau},  \nonumber \\ & \bar{\omega}=\sqrt{\frac{ m c \lambda_0^2 w_0^2}{4\alpha(A\!+\!B)}}\Delta\omega, \ \delta=\frac{2\sqrt{AB}}{A\!+\!B}, \ q_0=\frac{\pi w_0^2}{\lambda_0^2},
\end{align}
with $m$ as the mass of the particle and $\lambda_0$ as the beam wavelength. 
\begin{figure}[t!]
\includegraphics[width=0.48\textwidth,height=.20\textwidth,trim=.0cm 0.0cm 0cm 1.5cm,clip]{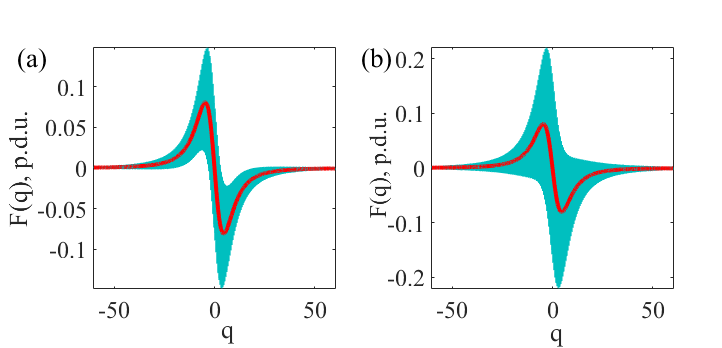}
\caption{Optical forces calculated from Eq~\eqref{potential}. Red lines show the static component of the force, while the shaded areas show the variation of the force per cycle;
$q_0=7.958,\ \bar{\omega}=0.342$. (a) $\delta=0.008$, (b) $\delta=0.011$.}
\label{fig2}
\end{figure}
In Fig.~\ref{fig2} we present plots of the force $F=-\partial V/\partial q$ versus $q$ for two sets of parameters.
\begin{figure}[t!]
\includegraphics[width=0.48\textwidth,height=.44\textwidth,trim=1.1cm 0.9cm 1.5cm 1.2cm,clip]{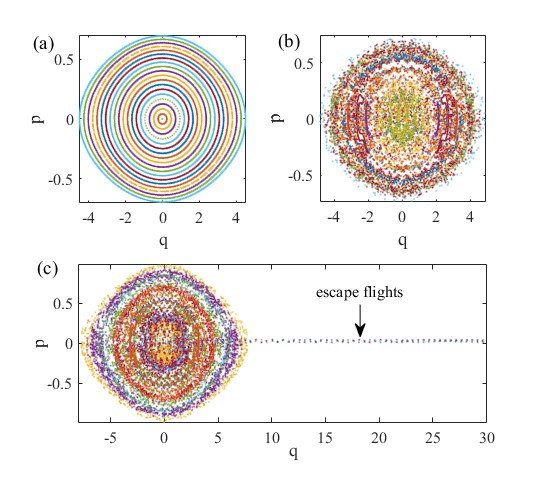}
\caption{Poincar\'e sections for Hamiltonian dynamics; $q_0=7.958,\ \bar{\omega}=0.342$.  (a) $\delta=0$, $\nu=1.16$; (b)  $\delta=0.008$, $\nu=1.59$; (c)  $\delta=0.011$, $\nu=1.55$ .}
\label{fig3}
\end{figure}

Equation~\eqref{Newton} is solved numerically to analyze the dynamics of the particles. The simulation results are shown in Fig.~\ref{fig3} in the form of Poincar\'e sections, i.e. by plotting the position and momentum of the particle at times that are multiples of $2\pi/\bar{\omega}$. Figure~\ref{fig3}~(a) shows the regime of regular dynamics for $\delta=0$ with motion on invariant tori. In Fig.~\ref{fig3}~(b) with $\delta=0.008$  almost all invariant tori are destroyed as the system enters a chaotic regime and the trajectories spread over phase space. The data visualized in Fig.~\ref{fig3}~(a) and Fig.~\ref{fig3}~(b) can be quantified by the correlation dimension $\nu$. The correlation dimension is calculated as the average over the correlation dimension of 20 trajectories sampled by choosing equidistant initial conditions with $p=0$, as seen in Fig.~\ref{fig3}~(a).  For Fig.~\ref{fig3}~(a) the averaged $\nu=1.16$, indicating regular motion, whereas for Fig.~\ref{fig3}~(a) $\nu=1.57$, which is a signature of chaos.  With a further increase of $\delta$ the motion remains chaotic near the center of the trap. However, there are trajectories in which the particles travel away from the trap centrum, see Fig.~\ref{fig3}~(c). Further on, we term such trajectories {\it escape flights}. The analytic solution for the escape flight can be found from the asymptotic expression for the optical force at large $q$,
\begin{equation}\label{far_force}
F(q,\tau)=-\frac{4\pi\delta}{1+\left( {q}/{q_0}\right)^2}\sin(4\pi q-\bar{\omega} \tau)+{\cal O}^3\left(\frac{q_0}{q}\right),
\end{equation}
as follows
\begin{equation}\label{z_wave}
q=\frac{\bar{\omega}}{4\pi}\tau+\frac{ n}{4}, \ n=\ldots,-2,0,2\ldots.
\end{equation}
Physically, Eq.~\eqref{z_wave} shows that the particle travels with the minima of the dynamic potential with phase velocity $\bar{\omega}/4\pi$ corresponding to wave-particle resonance. Note that the odd $n$ are dropped in Eq.~\eqref{z_wave} since they correspond to unstable solutions with the particle traveling at the maximum of the dynamic potential.

The particle escape can be qualitatively understood from
Fig.~\ref{fig2}. On can see in Fig.~\ref{fig2}~(a), which corresponds to Fig.~\ref{fig3}~(b), that there is a domains where the optical force is directed towards the trap center at any moment in time blocking the escape paths. On the contrary, in Fig.~\ref{fig2}~(c), which corresponds to Fig.~\ref{fig3}~(b), the optical force can be both attractive and repulsive for all $q$ allowing access to escape flights. Once the particle, having initially started in a chaotic trajectory, has reached the doorway Eq.~\eqref{z_wave} it switches to regular motion and leaves the trap. This scenario of a chaotic motion terminated in a regular trajectory is known as {\it transient chaos} \cite{tel1990transient}.  For $q_0=7.958$ and $\bar{\omega}=0.342$ the critical value of $\delta$ resulting in an open trap is found to be $\delta=0.0102$. It is worth mentioning that in the Hamiltonian case the escape flights can not be interpreted as attractors, since phase space volume contraction is forbidden by Liouville’s theorem, see Appendix A.

\begin{figure}[t!]
\includegraphics[width=0.49\textwidth,height=.22\textwidth,trim=0.8cm 1.7cm .8cm 2.5cm,clip]{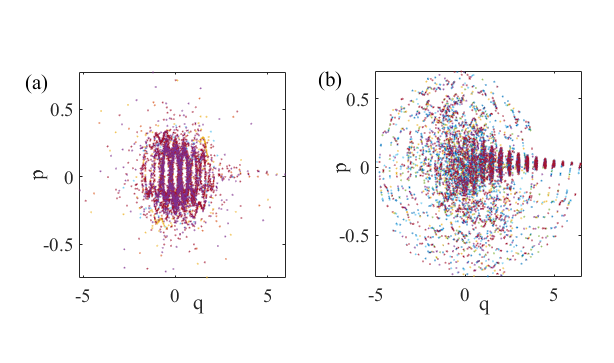}
\caption{Poincar\'e sections for dissipative dynamics; $q_0=7.958,\ \bar{\omega}=0.342$. (a)  $\delta=0.01$, $\kappa=0.0034$, $\nu=1.38$ (b)  $\delta=0.01$, $\kappa=0.0171$, $\nu=1.01$. }
\label{fig4}
\end{figure}
\begin{figure}[t!]
\includegraphics[width=0.45\textwidth,height=.35\textwidth,trim=3.5cm 0.9cm 3.5cm 1.2cm,clip]{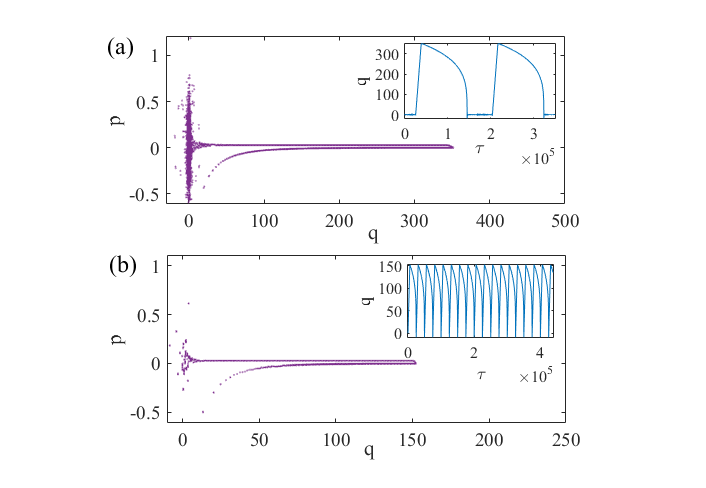}
\caption{Poincar\'e sections for dissipative escape trajectories. The insets show the distance against time; $q_0=7.958,\ \bar{\omega}=0.342$. (a)  $\delta=0.014$, $\kappa=0.0034$ (b)  $\delta=0.014$, $\kappa=0.0171$.}
\label{fig5}
\end{figure}

{\it Dissipative dynamics} -- The presence of the Stokes force modifies the equation of motion as follows
\begin{equation}\label{dis}
\frac{d^2q}{d\tau^2}=F(q,\tau)- \kappa \frac{dq}{d\tau},
\end{equation}
where
\begin{equation}\label{kappa}
\kappa=6\pi R \eta\sqrt{\frac{c  \lambda_0^2 w_0^2}{4\alpha m (A\!+\!B)}}
\end{equation}
with $\eta$ as viscosity.
Examples of localized motion are presented in Fig.~\ref{fig4} for two different values of $\kappa$. Fig.~\ref{fig4}~(a) demonstrates Poincar\'e sections for $\kappa=0.0034$ with the average correlation dimension $\nu=1.38$ which indicates coexistence of regular and chaotic motion. The correlation dimension diminishes $\nu=1.01$ with of increase of viscosity $\kappa=0.0171$, see Fig.~\ref{fig4}~(b).
A more interesting picture arises in an open trap that allows access to escape flights at  $\delta>0.0102$. One can in Fig.~\ref{fig5}~(a) that, similar to the Hamiltonian regime a particle escape is visible in the Poincar\'e section. Nonetheless, the escape flight is terminated, and the particle returns to the trap centrum at a slower speed, as one can see from the inset in Fig.~\ref{fig5}~(a). This scenario is repeated after some time within the trap. The termination of the escape flight can be explained by the asymptotic decrease of the dynamic potential which at a certain point can not counterbalance the Stokes force \cite{hanggi2005brownian}. So, the particle breaks away from the potential minimum, comes to a holt due to friction, and then returns to the trap under the action of the static potential, see Appendix B. With $\kappa=0.0171$ the escape distance becomes shorter due to the increase of the Stokes force. At the same time, the chaotic motion collapses within the trap and the particle almost immediately sets out for another escape flight. As a result, the particle performs periodic oscillations between the centrum and the periphery of the trap. The oscillation period can be assessed from the return flight time, Appendix B, which is the largest time scale in the oscillation dynamics, see the inset in Fig.~\ref{fig5}~(b). 

\begin{figure}[t!]
\includegraphics[width=0.45\textwidth,height=.22\textwidth,trim=.7cm 0.4cm 1.cm 1.5cm,clip]{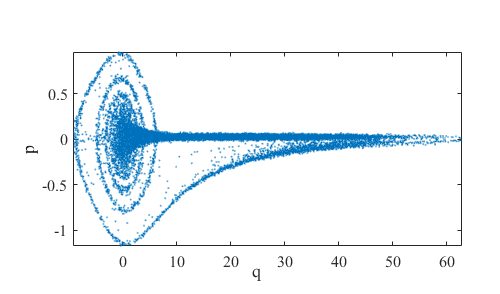}
\caption{Poincar\'e sections for dissipative escape trajectory parameters from Fig.~\ref{fig5}~(a) with account of the Brownian at $T=3 K$.}
\label{fig6}
\end{figure}

{\it Experimental feasibility} --  Let us now assess the values of parameters, which lead to the described effects in a realistic experimental set-up. First, we choose the radius of the particles $R=0.1~\mu{\rm m}$, the permittivity $\epsilon=12$, and the density of the particle material $\rho=2\cdot10^3~{\rm kg}/{\rm m}^3$. The geometric parameters of the beam are taken as $\lambda_0=1.55~\mu{\rm m} $ and $w_0=2.467~\mu{\rm m}$. An experimentally feasible power of the right-propagating beam is $A=1~{\rm mW}$, then the value of $B$ such as $B\ll A$ can be found from Eq.~\eqref{vars} as $B=\delta^2 A/4$. The dimensionless modulation frequency $\bar{\omega}=0.342$ leads to $\Delta\omega=3.16\cdot10^3{\rm~rad/s}$. Time in seconds can be calculated as $t=T_0\tau$ with $T_0=1.081\cdot10^{-4}{~\rm sec}$. According to Eq.~\eqref{kappa} for $\kappa=0.007$ we have $\eta=3\cdot10^{-10}{\rm~Pa\cdot~s} $ which is lower than the viscosity of air at room temperature and atmospheric pressure. First of all, we note that in high vacuum viscosity depends on air pressure \cite{Veijola1995}. We found that our estimates of viscosity is by orders of magnitude above the limit obtained in \cite{Dania2024} at low pressures. The choice of viscosity is also important because of the effect of Brownian motion as the viscosity enters the expression for the amplitude of the stochastic Brownian force. For observation of the particle escape it is necessary that the Brownian force be smaller than the force induced by the dynamic potential. The effect of the Brownian force can be suppressed by increase of the optical potential. To assess the effect of the Brownian force, we ran numerical simulation by solving the Langevin equation \cite{volpe2013simulation} at $A=100{\rm~mW}$ and  temperature $T=3~{\rm K}$ keeping the same values of the dimensionless parameters corresponding to Fig.~\ref{fig5}~(a). This choice of temperature is feasible in a view of the results on millikelvin cooling of optically trapped microparticles \cite{Li2011}. The simulation results are shown in Fig.~\ref{fig6}. One can see in Fig.~\ref{fig6} that, albeit the escape flight distance is shortened due to the Brownian motion, the particle still escapes to a distance $q=60$. 

\begin{figure}[t!]
\includegraphics[width=0.5\textwidth,height=.57\textwidth,trim=0.cm 0.cm 0.cm 0.5cm,clip]{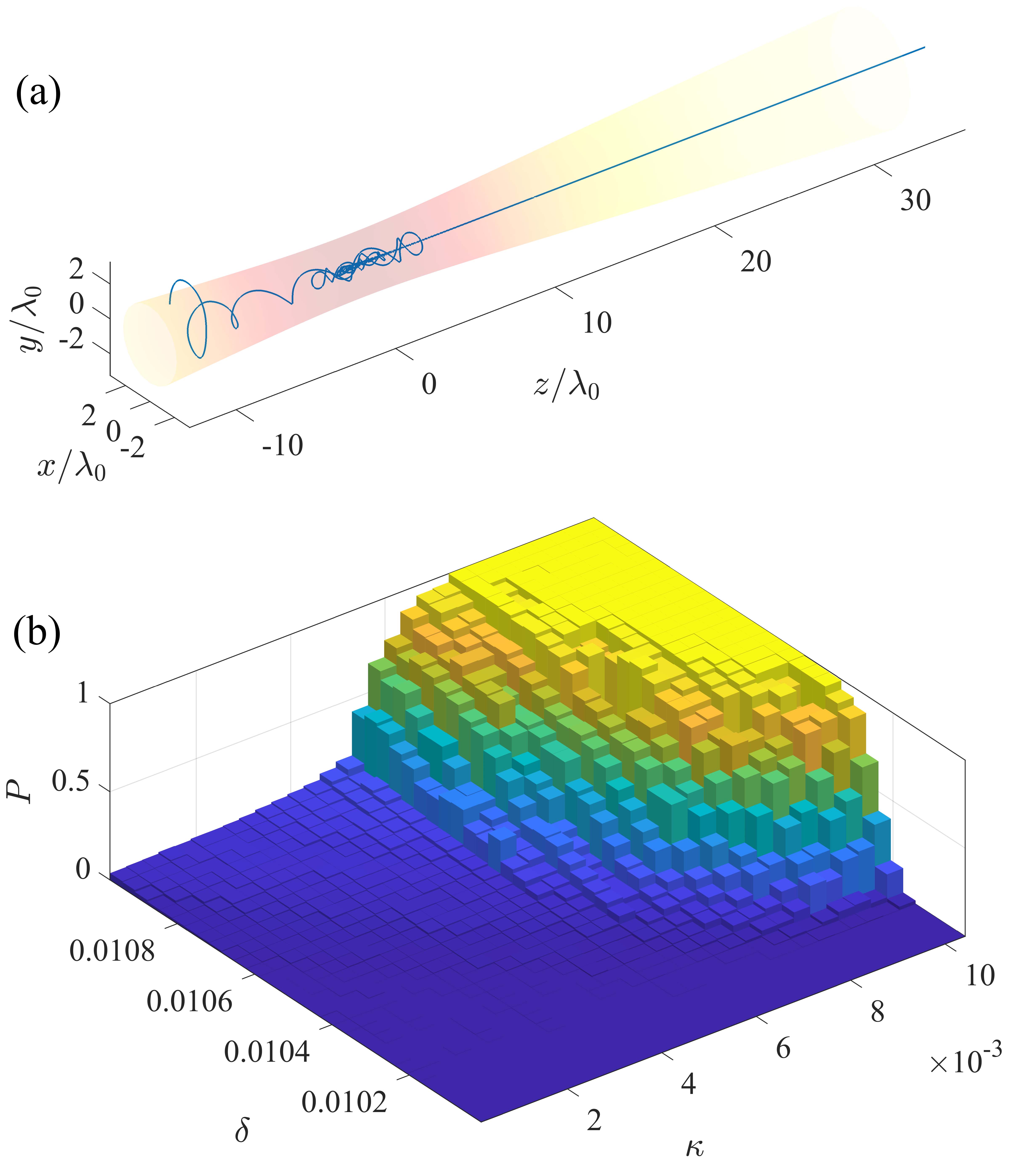}
\caption{Summary graphic data; $ q_0=7.958,\ \bar{\omega}=0.342$. (a)  Escape trajectory calculated in 3D space;  $\delta=0.02$, $\kappa=0.0273$. (b) Escape probability $P$ by distance $q=64.5$ at time $\tau=1.46\cdot 10^5$ versus  $\delta$ and $\kappa$.}
\label{fig7}
\end{figure}

{\it Summary and conclusion} -- We considered Rayleigh particles in a periodically modulated optical trap formed by two counter-propagating Gaussian beams. It is demonstrated that chaotic motion within the trap is terminated by regular escape flights in which the particle can travel hundreds micron away from the trap centrum. As a final note we point out that in realistic set-ups the initial conditions are set off from the beam axis. In Fig.~\ref{fig7}~(a) 
we present an example of such trajectory with the particle 
initially attracted to the trap centrum and then ejected into an escape flight. The threshold nature of particle escape, see Fig.~\ref{fig3}~(b) and Fig.~\ref{fig3}~(c), suggests possible applications for Rayleigh particle sorting. In Fig.~\ref{fig5}~(b) we show the escape probability as a function of two parameters, $\kappa$ and $\delta$. Since only $\kappa$ is dependent on the radius $R$, see Eq.~\eqref{kappa}, the escape is selective with respect to the size of the particle and, as such, can be potentially used for particle separation.
\newpage

The work was supported by Russian Science Foundation Grant No.
22-12-00070-P.

\bibliography{DNMaksimov}

\appendix

\section{Hamiltonian escape flights}
As it is pointed out in the main text, in the Hamiltonian case the escape flights cannot be seen as attractors because attractors are forbidden by Liouville’s theorem. This leads to an apparent contradiction since two-dimensional sets of initial conditions can not asymptotically evolve into the escape flights given by Eq.~(8) in the main text. To resolve this contradiction,  we start from the asymptotic equation of motion
\begin{equation}\label{Ham}
\frac{d^2q}{d\tau^2}=F(q,\tau),
\end{equation}
where
\begin{equation}\label{far_force}
F(q,\tau)=-\frac{4\pi\delta}{1+\left( {q}/{q_0}\right)^2}\sin(4\pi q-\bar{\omega} \tau)+{\cal O}^3\left(\frac{q_0}{q}\right).
\end{equation}
By writing the solution in a perturbative form
\begin{equation}
q=q_{\scs{(0)}}+q_{\scs{(1)}}, \ q_{\scs{(1)}}\ll q_{\scs{(0)}},
\end{equation}
where the zeroth order solution is the escape flight given by Eq.~(8) in the main text
\begin{equation}\label{unperturbed}
q_{\scs{(0)}}(\tau)=\frac{\bar{\omega}}{4\pi}\tau+\frac{ n}{4}, \ n=\ldots,-2,0,2,\ldots,
\end{equation}
one finds in the limit $\tau\rightarrow\infty$
\begin{equation}\label{difeq}
\frac{d^2q_{\scs{(1)}}}{d\tau^2}=-\frac{(4\pi)^3\delta q_0^2}{(\bar{\omega}\tau)^2} \sin\left(4\pi q_{\scs{(1)}} \right).
\end{equation}
The effective force on the right hand part of the above equation is directed in the opposite direction for small displacements $q_0$.
Thus, in addition to the translation motion at speed $\bar{\omega}/4\pi$ the particle can also perform oscillations about the minimum of the dynamic potential. The energy of these oscillations is dictated by the initial conditions on the transient chaotic trajectory. Since Eq.~\eqref{difeq} is a second order differential equations, the unperturbed escape flight is embedded into a two-parametric family of solutions. Thus, the particle escape does not contradict to Liouville’s theorem. For small oscillation the solution of Eq.~\eqref{difeq} can be found via the Taylor expansion of the sine, which leads to
\begin{equation}
\frac{d^2q_{\scs{(1)}}}{d\tau^2}+\frac{\alpha}{\tau^2}q_{\scs{(1)}}=0
\end{equation}
with
\begin{equation}
\alpha=\frac{(4\pi)^4q_0^2}{\bar{\omega}^2},
\end{equation}
and, consequently, to
\begin{equation}\label{sol}
q_{\scs{(1)}}(t)=\sqrt{t}\left[C_1\sin{(\beta \ln t)}+C_2\cos{(\beta \ln t)} \right],
\end{equation}
where
\begin{equation}
\beta=\frac{1}{2}\sqrt{4\alpha-1}.
\end{equation}
\begin{figure*}[ht!]
\includegraphics[width=0.8\textwidth,height=.34\textwidth,trim=.7cm 1.0cm 1.cm 2.2cm,clip]{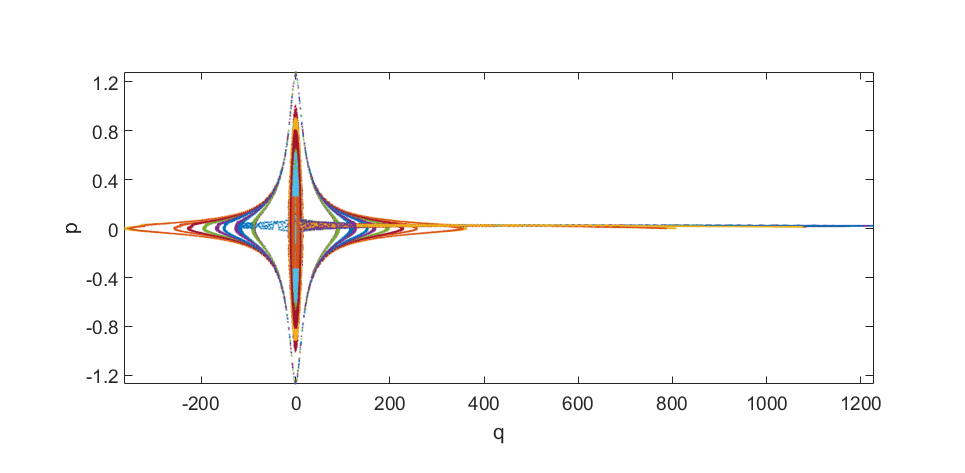}
\caption{Poincar\'e sections for Hamiltonian dynamics; $q_0=7.958,\ \bar{\omega}=0.342$, \ $\delta=0.014$.}
\label{figS1}
\end{figure*}
According to Eq.~\eqref{sol} the particle trapped by the traveling wave performs small oscillations with increasing amplitude and decreasing frequency. At large $\tau$ the oscillation amplitude becomes larger than the distance to the crest of the wave, and the escape flight is terminated. This conclusion also holds for the nonlinear Eq.~\eqref{difeq} since the anharmonicity leads to a decrease in the restoring force. In Fig.~\ref{figS1} we show the Hamiltonian escape flights calculated at large distances. One can see in Fig.~\ref{figS1} that the escape flights are terminated at some distance. The escape flight distance is determined by the parameters $C_1$ and $C_2$ in Eq.~\eqref{sol}, which, in turn, are dictated by the transient chaotic trajectory within the trap. All Hamiltonian escape flights are unstable with the exception of the core trajectory $C_1,C_2=0$.

\section{Dissipative escape flights}
We start by rewriting the equation of motion that controls the the dissipative dynamics
\begin{equation}\label{dis}
\frac{d^2q}{d\tau^2}=F(q,\tau)- \kappa \frac{dq}{d\tau}.
\end{equation}
First we note that, in contrast to the Hamiltonian case, in the dissipative case attractors are possible. Note, though, that Eq.~(8) in the main text is not the solution for the dissipative escape flight since it does not account of friction. In the dissipative case, the constant velocity escape solution is possible if the Stokes force is counterbalanced by the force induced the dynamic potential. Therefore in an escape flight the particle does not sit in the minimum of the dynamic potential but rather on its slope such that
\begin{equation}\label{disbal}
\kappa\frac{\bar{\omega}}{4\pi}=-\frac{4\pi\delta}{1+\left( {q_{\scs{(0)}}}/{q_0}\right)^2}\sin(q_{\scs{(1)}}).
\end{equation}
One can see from Eq.~\eqref{disbal} that $q_{\scs{(1)}}$ must be negative, so the particle is surfing on the frontal slope of the traveling wave dynamic potential. In the course of time, the dynamic potential is asymptotically decreasing. Therefore, once
\begin{equation}
q_{\scs{(1)}}=-\frac{\pi}{2}
\end{equation}
the dynamic potential can not overcome the Stokes force. The particle wave resonance is destroyed, and the motion comes to a halt. 
\begin{figure*}[t!]
\includegraphics[width=0.8\textwidth,height=.37\textwidth,trim=.7cm 2.5cm 1.cm 2.0cm,clip]{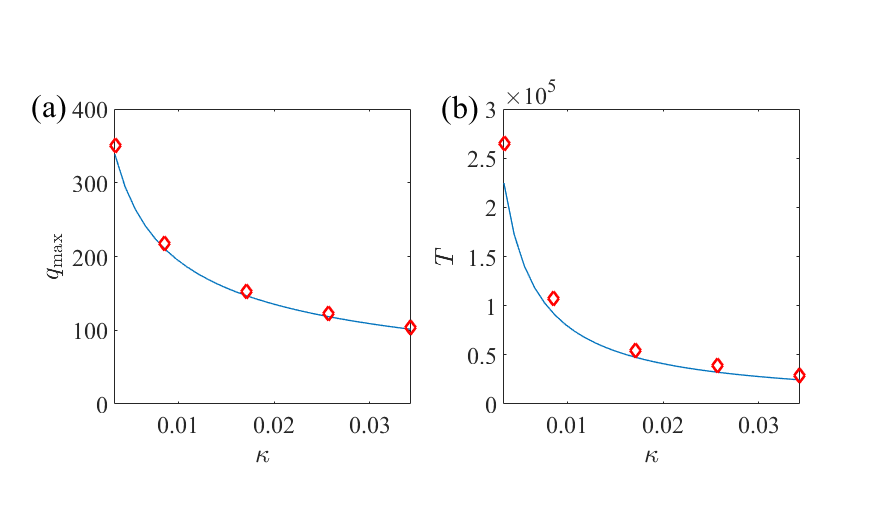}
\caption{Escape-return oscillations in the dissipative case $\delta=0.014, \bar{\omega}=0.342, q_0=7.958$. (a) The  escape distance versus $\kappa$. (b) The period versus $\kappa$. The blue lines show the semi-analytic solutions. The red diamonds depict the data obtain via direct numerical simulations.}
\label{figS}
\end{figure*}
As the motion slows down, and the dynamic potential oscillation period becomes much shorter than the travel time between the local minima of the potential. The action of the dynamic potential $V_d(q,t)$ is then averaged out over the characteristic time scale. Therefore, at the turning point the {\it surfer} becomes a {\it swimmer} \cite{hanggi2005brownian}, that  is only driven by the force produced by the static potential
\begin{equation}
V_s(q)=-\frac{1}{1+\left( {q}/{q_0}\right)^2}.
\end{equation}
This results in an overdamped return motion which is controlled by the equation below
\begin{equation}\label{over}
\frac{dq}{d\tau}=-\frac{1}{\kappa}\frac{\partial V_s(q)}{\partial q}.
\end{equation}
The return time $\tau_2$ can be estimated by solving Eq.~\eqref{over} numerically. Finally, the oscillation period in the escape trajectory which is composed from a synchronized escape flight and an overdamped return flight can be assessed as
\begin{equation}
T=\tau_1+\tau_2.
\end{equation}
The time $\tau_1$ can be neglected because the overdamped motion is much slower than the escape flight, see Fig.~(5) in the main text. To support our conclusions we applied Eq.~\eqref{disbal} and Eq.~\eqref{over} to estimate the escape distance $q_{{\rm max}}$ and the period of escape-return oscillations $T$ as dependent on $\kappa$. The results are shown in Fig.~\ref{figS}.

\end{document}